\documentclass[]{article}
\usepackage[T1]{fontenc}
\usepackage[utf8]{inputenc}
\usepackage[]{ismir} 
\renewcommand{\permission}{} 
\usepackage{amsmath,cite,url}
\usepackage{graphicx}
\usepackage{color}

\usepackage{booktabs}
\usepackage{amssymb}
\usepackage{pgfplots}
\usepackage[table]{xcolor}

\pgfplotsset{compat=1.18}
\title{Improving Music Source Separation with Diffusion and Consistency Refinement}

\multauthor
  {Tornike Karchkhadze$^1$$^{\dag}$
  \hspace{1cm} Mohammad Rasool Izadi$^2$$^{\dag}$ 
  \hspace{1cm} Shuo Zhang$^2$}
  {{\bf Shlomo Dubnov$^1$} \\
  $^1$ University of California San Diego, USA \\
  $^2$ Bose Corporation, USA \\
  }

\def\authorname{F. Author, S. Author, and T. Author}

\usepackage[bookmarks=false,pdfauthor={\authorname},pdfsubject={\pdfsubject},hidelinks]{hyperref}

\begin{document}

\maketitle

\def\thefootnote{\dag}
\footnotetext{Authors with equal contribution.}

\begin{abstract}
In this work, we propose an approach to music source separation that uses a generative diffusion model as a last-stage refinement on top of a deterministic separator, progressively enhancing the separated sources through iterative denoising. While the diffusion refinement yields measurable quality gains, it requires iterative steps at inference, increasing computational cost. To speed up the inference process, we apply consistency distillation, reducing inference to a single step while maintaining quality; with two or more steps, the distilled model even surpasses the diffusion-based approach. Crucially, our method is architecture-agnostic: we demonstrate state-of-the-art results when applied to both a custom U-Net-based separator on Slakh2100 and the state-of-the-art BS-RoFormer model on MUSDB18, showing that the refinement generalizes across backbone architectures. Sound examples are available at: \href{https://consistency-separation.github.io/}{https://consistency-separation.github.io/}.

\end{abstract}

\begin{figure*}[htbp]\centering
\includegraphics[width=0.95\textwidth]{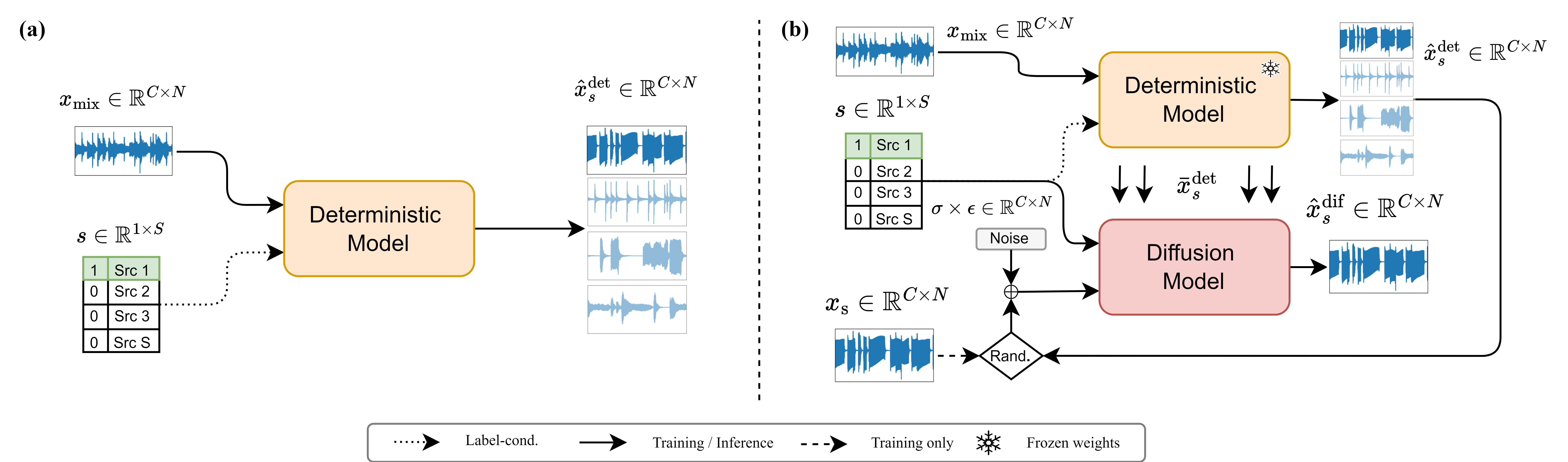}
     \caption{\small\textbf{Diagram illustrating our proposed method.} (a) The deterministic source separation model with optional one-hot instrument label conditioning. (b) Proposed system: the deterministic separator is augmented with a diffusion model conditioned on its intermediate features and instrument label, which further refines the separated audio through iterative denoising.}
    \label{fig:system}
\end{figure*}

\section{Introduction}
\label{introduction}

Source separation (SS) refers to the process of isolating individual sound sources from a mixture of audio signals. This task is crucial in various fields, including speech processing, noise reduction, music analysis, transcription, and more.
Music source separation (MSS), a subset of SS, is an inherently challenging problem. Instruments are often highly correlated, share overlapping frequency content due to harmonic relationships, and instruments of the same family share timbral characteristics that make them difficult to distinguish~\cite{cano2019musical, watcharasupat2025separate}. These factors often lead to incomplete target source reconstruction, residual source leakage and reconstruction artifacts in the separated output.

Machine learning has driven substantial progress in SS, with two primary paradigms emerging: deterministic and generative. The first involves deterministic discriminative models \cite{choi2021lasaft, defossez2021hybrid, lluis2019end, guso2022loss, luo2019conv, defossez2019music, takashi2018mmdenselstm, rouard2023hybrid, luo2023music, lu2024music, wang2024melroformer}, which typically use mixtures for conditioning and learn to regressively derive one or more sources from the mixture. On the other hand, generative models~\cite{subakan2018generative, kong2019single, narayanaswamy2020unsupervised, jayaram2021parallel, postolache2023, postolache2023adversarial, kavalerov2019universal, scheibler2025flowmatching, wisdom2020unsupervised, scheibler2023diffusion, Huang2023DAVIS, yu2023zeroshot, plaja2022adiffusion, wen2025guidesep, zhu2022music, karchkhadze2024simultaneous, mariani2024multisource} learn a distribution over sources and generate by sampling from it conditioned on the mixture. Hybrid approaches combining both paradigms have also been explored~\cite{manilow2022improving, SchafferCMMSP22, lutati2023separate, hirano2023diffusion, LemercierRWG23, lemercier2023wind, shi2024diffusion}. 
Today, state-of-the-art performance is still dominated by deterministic models, yet they remain limited by the inherent challenges of separation, that regression-based objectives struggle to fully resolve. While generative models underperform deterministic ones, their ability to learn a prior distribution suggests that intelligently combining the two paradigms may benefit separation performance, addressing what deterministic models leave unresolved~\cite{lutati2023separate, araki2025years}.

To address the limitations described above, we introduce a denoising score-matching diffusion model~\cite{song2019generativemodeling, song2021scorebased} as a last-stage generative refinement on top of a pretrained deterministic separator. Since generative models can model the distribution of clean sources and synthesize data from scratch, we hypothesize that incorporating a generative component would help the model reconstruct missing information and further improve the quality of MSS.
However, the iterative sampling procedure of diffusion models introduces additional inference latency. To mitigate this, we apply Consistency Distillation (CD)~\cite{song2023consistency, Kim24CTM}, reducing inference to a single step.

We first build and train our own time-domain U-Net-based separator on the Slakh2100~\cite{manilow2019cutting} dataset. Applying the diffusion and CD refinement on top of this model yields significant improvements in objective separation metrics, establishing a new state-of-the-art on Slakh2100 compared to Demucs~\cite{defossez2019music}, Demucs+Gibbs~\cite{manilow2022improving}, and MSDM~\cite{mariani2024multisource}. Furthermore, our CD model achieves accelerated single-step denoising without loss of quality, and with two or more steps surpasses the diffusion-based approach.
To demonstrate the model-agnostic nature of our approach and benchmark against state-of-the-art, we adopt BS-RoFormer~\cite{lu2024music} --- the best-performing publicly available separator --- as the deterministic backbone on MUSDB18~\cite{MUSDB18}. Following our method, a second BS-RoFormer is built to serve as the diffusion model and trained on top of the deterministic one. This yields consistent improvements in objective separation metrics over the strong BS-RoFormer baseline, setting a new state-of-the-art on MUSDB18. Finally, the consistency-distilled model achieves equivalent performance to the diffusion model in a single step, and surpasses it with two and more steps.

As part of our commitment to reproducibility and open science, the code and checkpoints are publicly available\footnote{\href{https://github.com/Russell-Izadi-Bose/DiCoSe}{https://github.com/Russell-Izadi-Bose/DiCoSe}}.

\section{Related Work}\label{related}

Deterministic models have long dominated SS, with early models focused on spectrogram-based approaches, initially for speech~\cite{classif_dnn_separation} and later for music~\cite{uhlich2015deep, uhlich2017improving, liu2018denoising, sony_densenet, multichannel_deep, takahashi2020d3net, spleeter2020}. Subsequent work shifted interest to waveform-domain approaches, first in speech~\cite{lluis2019end, luo2019conv}, followed by MSS models such as Wave-U-Net~\cite{wavunet} and Demucs~\cite{defossez2019music}, the latter being a milestone work extending the U-Net architecture with bidirectional LSTM layers. Later research explored hybrid time-frequency architectures combining both domains — Hybrid Demucs~\cite{defossez2021hybrid} and HT Demucs~\cite{rouard2023hybrid}. The current state of the art has shifted to purely frequency-domain models that, unlike earlier magnitude-spectrogram approaches, operate on complex STFT representations. A key innovation is band-splitting (BS) — partitioning the spectrum into frequency subbands processed independently: Band-Split RNN~\cite{luo2023music} uses recurrent networks per subband, while BS-RoFormer~\cite{lu2024music} and Mel-RoFormer~\cite{wang2024melroformer} replace recurrent layers with Rotary-Embedding Transformers, with BS-RoFormer achieving the best published results on MUSDB18.

On the other hand, purely generative approaches to SS have also been explored~\cite{subakan2018generative, kong2019single, narayanaswamy2020unsupervised, jayaram2021parallel, postolache2023, postolache2023adversarial, kavalerov2019universal, wisdom2020unsupervised, scheibler2023diffusion, scheibler2025flowmatching}, including GAN-based, flow-based, and diffusion-based models. Analogous approaches have been proposed specifically for MSS~\cite{zhu2022music, karchkhadze2024simultaneous, yu2023zeroshot, plaja2022adiffusion, wen2025guidesep}, however, these approaches typically score lower than the deterministic models on reference-based objective metrics such as SDR. MSDM~\cite{mariani2024multisource}, currently the strongest purely generative approach on standard objective metrics, introduced a score-matching diffusion model for simultaneous MSS and generation in the waveform domain. Since our method uses consistency distillation, we note that consistency models have previously been explored for audio generation and compression~\cite{saito2024soundctm, fei2024musiccm, pasini2024music2latent}, but not, to our knowledge, for MSS.

Several approaches have combined discriminative and generative components for SS across both music and speech. In MSS, Demucs+Gibbs~\cite{manilow2022improving}, building on Demucs, introduced an iterative Gibbs-sampling refinement that enforces mixture consistency across sources. MSG~\cite{SchafferCMMSP22} layers a GAN-based generative refinement stage on top of Demucs, achieving perceptual improvements. In the speech domain, Diffiner~\cite{hirano2023diffusion} applies a diffusion-based refiner as a post-processor on top of any existing speech separator, improving perceptual quality without retraining the preceding model. Closely related to our work, Separate and Diffuse~\cite{lutati2023separate} applies a pretrained diffusion vocoder as post-processing on top of a deterministic speech separator, but operates on Mel-Spectrograms, requiring phase reconstruction and a separate learned combining network — neither of which our  approach requires.

More tangentially related to our work, in the music domain, BABE~\cite{moliner2023babe} and BABE-2~\cite{moliner2024babe2} apply diffusion posterior sampling for blind audio bandwidth extension and music restoration, targeting enhancement of a single degraded recording rather than source separation. Similar hybrid approaches have been explored for speech enhancement~\cite{LemercierRWG23, lemercier2023wind, shi2024diffusion}, though, different from our work, these operate in the spectrogram domain and mostly target only perceptual quality improvements.

\section{Method}

Let \( x_{\text{mix}} \) represent a time-domain audio mixture containing \( S \) individual sources \( x_s \in \mathbb{R}^{C \times N} \), where \( C \) is the number of channels, \( N \) is the number of audio samples, and \( s \in \{1, \dots, S\} \) identifies each source. The mixture is defined as \( x_{\text{mix}} = \sum_{s=1}^S x_s \). The SS problem is to recover each source from \( x_{\text{mix}} \) so that the true sources \( x_s \) and their estimates \( \hat{x}_s \) are as close as possible.

\subsection{Deterministic Model}

Let \( f_{\theta} \) denote a deterministic source separator, shown in the left side of Fig.~\ref{fig:system}. For label-conditioned models, \( f_{\theta} \) takes both the mixture \( x_{\text{mix}} \) and a one-hot source label \( s \in \mathbb{R}^{1 \times S} \) as input, applied once per source by iterating over all \( s \in \{1, \dots, S\} \). For models that separate all sources simultaneously, \( f_{\theta} \) takes only \( x_{\text{mix}} \) and produces all estimates jointly. We denote the estimate for source \( s \) as \( \hat{x}_s^{\text{det}} \) and train \( f_{\theta} \) with:
\begin{equation}
    \label{eq:loss_det}
    \mathcal{L}(\theta) = \mathbb{E}_{s, x_{\text{mix}}} \left[ \ell\!\left(x_s,\, \hat{x}_s^{\text{det}}\right) \right],
\end{equation}
where \( \ell(\cdot, \cdot) \) is a model-specific distance-based loss (e.g., MSE, L1, multi-resolution spectral, or a combination).

\subsection{Diffusion Model}
We leverage a diffusion generative model~\cite{dickstein2015deepunsupervised, ho2020denoising}, more specifically a Denoising Score Matching (DSM)~\cite{song2019generativemodeling, song2021scorebased} formulation, which learns to denoise a signal by estimating the gradient of the data distribution. After training \(f_{\theta}\), we freeze its parameters and integrate it into a larger system by adding a diffusion model \(g_{\phi}\), as depicted in the right side of Fig.~\ref{fig:system}.
The diffusion model \(g_{\phi}\) takes as input: (1) a noisy version of the source signal \(\tilde{x}_s\), (2) the source label \(s\), and (3) the intermediate features \(\bar{x}_s^{\text{det}}\) extracted by the frozen deterministic model.
During training, the input \(\tilde{x}_s\) is randomly drawn from the deterministic estimate or the clean source with probability \(p\), encouraging robustness to imperfect separation:
\begin{equation}\label{eq:bernoulli_mix}
    \tilde{x}_s = (1-b)\,\hat{x}_s^{\text{det}} + b\, x_s, \quad b \sim \text{Bernoulli}(p).
\end{equation}
We train \(g_{\phi}\) with the DSM objective following EDM~\cite{karras2022elucidating}:
\begin{equation}\label{eq:dsm_loss_ours}
    \mathcal{L}_{\text{DSM}}(\phi) = \mathbb{E}_{s, x_{\text{mix}}, \sigma} \| x_s - g_{\phi}(\tilde{x}_s + \sigma \epsilon, s, \sigma, \bar{x}_s^{\text{det}}) \|_2^2,
\end{equation}
where \( \epsilon \sim \mathcal{N}(0, I)\) is Gaussian noise and \(\sigma\) is the noise level sampled from a log-normal distribution.

The inference of \(g_{\phi}\) is an iterative process over \(T\) discrete steps, where \(\sigma_t\) decreases from \(\sigma_{\text{max}}\) to \(\sigma_{\text{min}}\).
Crucially, the process is initialized from the deterministic estimate rather than pure noise: \(\hat{x}_{s,T}^{\text{dif}} = \hat{x}_s^{\text{det}} + \sigma_{\text{max}}\,\epsilon\).
Each subsequent step refines the estimate:
\begin{equation}\label{ode_solver}
    \hat{x}_{s, t-1}^{\text{dif}} = \texttt{Solver}_{1}(\hat{x}_{s, t}^{\text{dif}}, s, \sigma_t, \bar{x}_s^{\text{det}}; g_{\phi}),
\end{equation}
where \(\texttt{Solver}_k(\dots; g_{\phi})\) denotes \(k\) steps of a numerical solver using \(g_{\phi}\) for denoising, until reaching the final estimate \(\hat{x}_{s, 0}^{\text{dif}}\).


\subsection{Consistency Model}

To mitigate the latency introduced by the iterative sampling of the diffusion model in the inference process and make 1-2 steps generation possible, we adopt Consistency Distillation (CD) inspired by methods shown in \cite{song2023consistency, Kim24CTM}. 
In this approach, our consistency model \( g_{\omega} \) is designed as an exact replica of the diffusion model and is trained using a pretrained diffusion model \(g_{\phi}\) as a teacher. Requiring inference of diffusion teacher model, CD is designed as a discrete process with $t \in [1, T]$, where $T$ denotes a total number of steps.
We adopt a CD procedure~\cite{song2023consistency} augmented with two key elements from CTM~\cite{Kim24CTM}: multistep numerical solvers for the teacher instead of single-step, and an auxiliary DSM loss. Unlike CTM, we distill directly toward the clean estimate rather than an intermediate point along the diffusion trajectory. The teacher produces a less noisy target by running \(h\) solver steps:
\begin{equation}
    \hat{x}_{s, t-h}^{\text{dif}} = \texttt{Solver}_h( \tilde{x}_{s} + \sigma_{t} \epsilon, s, \sigma_{t}, \bar{x}_s^{\text{det}}; g_{\phi}),
\end{equation}
where \( h \in [1, t]\) is the number of solver steps used in the distillation process, and \(\tilde{x}_s\) is the Bernoulli-mixed input from Eq.~\eqref{eq:bernoulli_mix}.
This prediction is then used to calculate the target in the final CD loss:
\begin{equation}
\resizebox{0.7\linewidth}{!}{$\displaystyle
\begin{aligned}
    \mathcal{L}_{\text{CD}}({\omega}) = \mathbb{E}_{t, h} \| \underbrace{g_{\texttt{sg}(\omega)}(\hat{x}_{s, t-h}^{\text{dif}}, s, \sigma_{t-h} , \bar{x}_s^{\text{det}})}_{\textit{target}} - \\
    \underbrace{g_{{\omega}}(\tilde{x}_{s} + \epsilon \sigma_t, s, \sigma_t , \bar{x}_s^{\text{det}})}_{\textit{prediction}}\|_2^2
\end{aligned}
$}
\end{equation}
where \(\texttt{sg}(\omega)\) denotes the stop-gradient running EMA (Exponential Moving Average) of \(\omega\) during optimization, updated as 
\(\texttt{sg}(\omega) \leftarrow \texttt{stopgrad}(\mu \texttt{sg}(\omega) + (1 - \mu) \omega)\), with \(\mu\) denoting EMA update rate.



The final training objective combines \(\mathcal{L}_{\text{CD}}\) with the DSM loss from Eqn~\eqref{eq:dsm_loss_ours}:
\begin{equation} \label{eq:our_final}
\mathcal{L}({\omega}) = \mathcal{L}_{\text{CD}}({\omega}) + \lambda_{\text{DSM}}\mathcal{L}_{\text{DSM}} (\omega),
\end{equation}
where  $\lambda_{\text{DSM}}$ is a balancing term between two losses. 


\section{Experimental setup}\label{sec:experimental}

We apply our method to two deterministic backbones: a custom time-domain U-Net and a pre-trained BS-RoFormer~\cite{lu2024music}. We refer to these as the \textbf{U-Net experiments} and \textbf{BS-RoFormer experiments} throughout.

\subsection{Dataset and baselines}
\label{sec:baslines}

For our U-Net experiments, we use Slakh2100~\cite{manilow2019cutting}, a synthetically generated benchmark of 2100 multi-track recordings covering Bass, Drums, Guitar, and Piano (1500/375/225 train/val/test split). Its large scale makes it well-suited for the data-hungry diffusion model. We compare against the generative MSDM~\cite{mariani2024multisource} and the hybrid Demucs+Gibbs~\cite{manilow2022improving}, following the exact MSDM setup: 22kHz mono ($C=1$) audio with segment length \(N=262144\) samples (\(\approx 11.9\) sec.) for direct comparability.

For our BS-RoFormer experiments, we use MUSDB18-HQ~\cite{MUSDB18}, comprising 100 training and 50 test tracks across Bass, Drums, Vocals, and Other. We adopt the augmentation pipeline of BS-RoFormer~\cite{lu2024music}, including loudness variation, source mixup, pitch shifting, parametric EQ, tanh distortion, and polarity/channel augmentations, with a segment length of \(N=485100\) samples (11 sec., 44.1kHz, $C=2$). MUSDB18 is a long-standing standard benchmark in MSS, and we compare against the strongest deterministic separation models evaluated on it: Demucs~\cite{defossez2019music}, Hybrid Demucs~\cite{defossez2021hybrid}, HT Demucs~\cite{rouard2023hybrid}, BS RNN~\cite{luo2023music}, and BS-RoFormer. 

Across both experiments, we compare against MSG~\cite{SchafferCMMSP22} — our closest conceptual competitor, which similarly combines a Demucs deterministic separator with a generative GAN refinement for MSS. 

\subsection{Model Architectures and Training}
\label{sec:arch}

\subsubsection{U-Net}
\label{sec:unet_arch}

In the U-Net experiments, we build and train a custom 1D waveform-domain U-Net operating on mono audio, following the architecture of Moûsai~\cite{schneider2023mousai} and MSDM~\cite{mariani2024multisource}: six encoding levels with ResNet blocks and multi-head self-attention at the three deepest levels (8 heads, 128 features per head), 256 base channels, and downsampling factors of \([4,4,4,2,2,2]\) across levels. We further adapt the architecture with one-hot instrument label conditioning (\(S=4\)), following the success of this approach in SS~\cite{meseguer2019conditioned, choi2021lasaft}. It is trained with time-domain MSE loss, Adam at \(1 \times 10^{-4}\) for 170 epochs.

The Diffusion model shares the same U-Net backbone, extended with a diffusion time embedding that is projected and applied as FiLM~\cite{perez2018film} scale-shift conditioning at every ResNet block. Intermediate features \(\bar{x}_s^{\text{det}}\) from the frozen Deterministic model — extracted at every U-Net layer, matching the dimensions of the corresponding Diffusion model layers — are directly added to the Diffusion activations at each layer (see Fig.~\ref{fig:system}). Bernoulli mixing uses \(p=0\) (always clean source as diffusion input). Both training and inference follow the EDM framework~\cite{karras2022elucidating}: training uses \(\ln(\sigma) \sim \mathcal{N}(-3.0,\, 1.0^2)\), \(\sigma_\text{data}=0.2\), Adam at \(1 \times 10^{-4}\) for 280 epochs; inference uses the stochastic sampler with $T$ solver steps, $R$ correction steps, and stochasticity $S_{\text{churn}}$ — values reported with results.

The Consistency model is initialized from the pre-trained Diffusion model. We fix \(T=18\), \(\mu=0.999\), and use up to \(h \leq 17\) Heun solver steps. Training balances the CD objective with an auxiliary DSM loss (\(\lambda=0.7\)) as in Eqn~\eqref{eq:our_final}, using RAdam~\cite{liu2019variance} at \(1 \times 10^{-5}\) for 50 epochs.

\subsubsection{BS-RoFormer}
\label{sec:bs_rf_arch}

In our BS-RoFormer experiments, we use the publicly available pre-trained BS-RoFormer checkpoint\footnote{\url{https://github.com/ZFTurbo/Music-Source-Separation-Training}} as the Deterministic backbone. BS-RoFormer is a frequency-domain model (dim=384, depth=8) consisting of a shared band-split projection followed by a stack of time and frequency transformers that process the stereo mixture, whose output is then passed to 4 separate per-stem mask estimator heads to reconstruct each source. It is trained with a combined L1 and multi-resolution STFT loss.

For the Diffusion model, we extend the BS-RoFormer while preserving its full architecture so that the Diffusion model is identical in size and structure to the Deterministic model, in line with our method's design principle. We make each forward pass stem-specific by introducing a learned stem embedding layer that is summed with the diffusion time embedding and applied as FiLM scale-shift conditioning at every transformer layer, while the 4 separate mask estimator heads are preserved as in the original BS-RoFormer. Deterministic features are injected via zero-initialized adapter layers added to the corresponding Diffusion model stages.
Training and inference follow EDM framework with \(\ln(\sigma) \sim \mathcal{N}(-3.0,\, 1.0^2)\) and \(\sigma_\text{data}=0.06\). During training, a single randomly selected stem is processed per step, with the model seeing the clean source or the Deterministic extracted stem with equal probability (\(p=0.5\) Bernoulli mixing), using Adam at \(1 \times 10^{-4}\) for 80 epochs (4000 steps each); at inference all stems are processed sequentially.

The Consistency model is initialized from the Diffusion checkpoint. We fix \(T=10\), \(\mu=0.999\), and use up to \(h \leq 9\) Heun solver steps. Training balances the CD objective with an auxiliary DSM loss (\(\lambda=0.7\)), with Bernoulli mixing \(p=0.5\), using RAdam at \(1 \times 10^{-5}\) for TBD epochs (4000 steps each).


\subsection{Evaluation Metrics}
\label{eval}

We report the same two metrics for both experimental tracks. The scale-invariant Signal-to-Distortion improvement (SI-SDR\(_\text{I}\))~\cite{roux2019sdr} is computed using a sliding window of 4 seconds with 2-second overlap, filtering out silent chunks and those with only a single non-silent source; for Slakh2100 this follows the exact procedure of MSDM and Demucs+Gibbs. We also report Signal-to-Distortion Ratio (SDR), following the evaluation protocol from ~\cite{stoter20182018} using the \textit{museval} Python package, which computes the median SDR over 1-second chunks across tracks.


\begin{table}[t]
\centering
\resizebox{\linewidth}{!}{%
\begin{tabular}{lccccc}
    \toprule
    \multicolumn{6}{c}{\textbf{SI-SDR\(_\text{I}\)$^\dagger$ (dB) $\uparrow$}} \\
    \midrule
    \textbf{Model}  & \textbf{Bass} & \textbf{Drums} & \textbf{Guitar} & \textbf{Piano} & \textbf{All} \\ 
    \midrule
    Demucs (s=0) \cite{defossez2019music} (rep. in \cite{manilow2022improving})  & 15.77 & 19.44 & 15.30 & 13.92 & 16.11 \\
    Demucs (s=0) + Gibbs (512) \cite{manilow2022improving} & 17.16 & 19.61 & 17.82 & 16.32 & 17.73 \\ 
    ISDM (Dirac) \cite{mariani2024multisource} & 19.36 & 20.90 & 14.70 & 14.13 & 17.27 \\ 
    MSDM (Dirac) \cite{mariani2024multisource} & 17.12 & 18.68 & 15.38 & 14.73 & 16.48 \\ 
    \midrule
    U-Net Det.  & 19.99 & 20.80 & 23.74 & 20.76 & 21.32 \\ 
    U-Net Det. $\times 2$ & 20.93 & 20.97 & 23.56 & 21.14 & 21.65  \\
    \midrule
    U-Net Det. + MSG  & 19.52 & 20.55 & 23.06 & 20.20 & 20.83 \\
    \midrule
    U-Net + Diff. \scriptsize{($T \times R=2\times 2, \sigma_{\text{max}}=0.01$)} & 20.37 & 21.85 & 26.33 & 23.47 & 23.00 \\ 
    \midrule
    U-Net + CD \scriptsize{($T=1, \sigma_{\text{max}}=0.012$)} & 20.99 & 21.91 & 26.10 & 22.73 & 22.93  \\
    U-Net + CD \scriptsize{($T=2, \sigma_{\text{max}}=0.25$)} & 22.02 & 22.24 & 27.48 & 23.37 & 23.78  \\ 
    U-Net + CD \scriptsize{($T=4, \sigma_{\text{max}}=0.25$)} & \textbf{22.50} & \textbf{22.42} & \textbf{28.05} & \textbf{23.81} & \textbf{24.20}  \\ 
    \bottomrule
\end{tabular}
}


\resizebox{\linewidth}{!}{%
\begin{tabular}{lccccc}
    \toprule
    \multicolumn{6}{c}{\textbf{SDR (dB) $\uparrow$}} \\
    \midrule
    \textbf{Model}  & \textbf{Bass} & \textbf{Drums} & \textbf{Guitar} & \textbf{Piano} & \textbf{All} \\ 
    \midrule
    U-Net Det.  & 14.96 & 16.21 & 6.07 & 6.32 & 10.89 \\
    U-Net Det. $\times 2$ & 15.12 & 16.40 & 5.87 & 6.40 & 10.95 \\
    \midrule
    U-Net Det. + MSG  & 14.59 & 15.70 & 6.13 & 6.27 & 10.67 \\
    \midrule
    U-Net + Diff. \scriptsize{($T \times R=2\times 2, \sigma_{\text{max}}=0.01$)} & 15.36 & 16.52 & 6.59 & 6.85 & 11.34 \\
    \midrule
    U-Net + CD \scriptsize{($T=1, \sigma_{\text{max}}=0.012$)} & 15.76 & 16.92 & 6.36 & 6.65 & 11.42 \\
    U-Net + CD \scriptsize{($T=2, \sigma_{\text{max}}=0.25$)}  & 15.78 & 17.19 & 6.71 & 7.23 & 11.73 \\
    U-Net + CD \scriptsize{($T=4, \sigma_{\text{max}}=0.25$)}  & \textbf{16.13} & \textbf{17.38} & \textbf{6.89} & \textbf{7.39}  & \textbf{11.95} \\       
       
    \bottomrule
\end{tabular}
}

\caption{\small\textbf{U-Net Experiments: Source Separation Results on Slakh2100.} The upper section presents \textbf{SI-SDR\(_\text{I}\)} results, while the lower section presents \textbf{SDR} results, comparing our U-Net deterministic, diffusion, and CD models against baselines. $^\dagger$Follows the MSDM evaluation pipeline, including silent stems.} 
\label{tab:results_sep}
\end{table}

\section{Results and Discussion}

\subsection{U-Net on Slakh2100}

Table~\ref{tab:results_sep} reports SI-SDR\(_\text{I}\) and SDR results on Slakh2100 test set, comparing our U-Net experiment models against the baselines. The results show the following: 

\textbf{Deterministic model.} Our Deterministic model outperforms all baselines. Against the generative baselines (ISDM and MSDM), this is largely expected — deterministic models generally have an advantage in objective metrics over generative ones. The more informative comparison is with Demucs (s=0, no shift trick)~\cite{defossez2019music} retrained on Slakh by~\cite{manilow2022improving}, which shares the same paradigm: a waveform-domain U-Net encoder-decoder with skip connections. We attribute our advantage to three architectural differences: our model uses multi-head self-attention at the bottleneck rather than a bidirectional LSTM, one-hot label conditioning to extract one stem at a time rather than all 4 simultaneously, and a substantially larger model size ($\approx$10$\times$, 405M vs.\ 40M as reported by~\cite{mariani2024multisource}).

\textbf{Diffusion model.} Adding the Diffusion model further improves separation quality, seen in both SI-SDR\(_\text{I}\) (+1.7 dB) and SDR (+0.45 dB). We found \(T \times R = 2 \times 2\), \(\sigma_{\text{max}} = 0.01\), and \(S_{\text{churn}} = 20.0\) to perform best for our Diffusion model. To verify the gain is not merely due to stacking a second model and increasing of parameter count, we train a second Deterministic model on top of the frozen first (Det.$\times$2) — same two-model architecture as Det.+Diff., but without noise injection or iterative refinement. As seen in table, U-Net Det.$\times$2 outperforms the single U-Net Det.\ but remains below the Diffusion model, confirming the gains stem from the generative component.

Contextualising against other hybrid deterministic–generative methods: Demucs+Gibbs~\cite{manilow2022improving} achieves a +1.6 dB SI-SDR\(_\text{I}\) improvement over Demucs, slightly below our Diffusion step of +1.7 dB. To further compare with MSG~\cite{SchafferCMMSP22}, we integrate it into our pipeline and train it ourselves on top of our U-Net Det.\ model, as it was not trained or evaluated on Slakh2100. Despite clear perceptual improvements, we did not observe any gain in either SI-SDR\(_\text{I}\) or SDR — consistent with the authors’ own focus on perceptual rather than objective quality.  

\begin{table}[t]
\centering

\resizebox{\linewidth}{!}{%
    \begin{tabular}{lccccc}
        \toprule
        \multicolumn{6}{c}{\textbf{SI-SDR\(_\text{I}\) (dB) $\uparrow$}} \\
        \midrule
        \textbf{Model}  & \textbf{Bass} & \textbf{Drums} & \textbf{Other} & \textbf{Vocals} & \textbf{All} \\ 
        \midrule
        Demucs (s=10)  & 11.12 & 10.93 & 8.68 & 12.01 & 10.69 \\

        Demucs (s=10) + MSG & 10.23 & 8.79 & -- & 11.34 & 10.12 \\
        
        \midrule
        BS-RF Det. & 14.93 & 15.95 & 12.47 & 17.05 & 15.10 \\
        BS-RF Det. $\times 2$ & 14.90 & 15.74 & 12.42 & 16.99 & 15.01 \\
        \midrule
        BS-RF + Diff. \scriptsize{($T=3, \sigma_{\text{max}}=0.002$)} & 15.73 & 16.45 & 12.75 & 17.38 & 15.58 \\
        \midrule
        BS-RF + CD \scriptsize{($T=1, \sigma_{\text{max}}=0.004$)} & 15.92 & 16.53 & \textbf{12.83} & \textbf{17.54} & \textbf{15.71} \\
        BS-RF + CD \scriptsize{($T=2, \sigma_{\text{max}}=0.003$)} & \textbf{15.99} & \textbf{16.56} & 12.81 & 17.50 & \textbf{15.71} \\

        \bottomrule
    \end{tabular}
}

\resizebox{\linewidth}{!}{%
    \begin{tabular}{lccccc}
        \toprule
        \multicolumn{6}{c}{\textbf{SDR (dB) $\uparrow$}} \\
        \midrule
        \textbf{Model}  & \textbf{Bass} & \textbf{Drums} & \textbf{Other} & \textbf{Vocals} & \textbf{All} \\ 
        \midrule
        Demucs (s=10)~\cite{defossez2019music} & 7.01 & 6.86 & 4.42 &  6.84 & 6.28 \\
        HT Demucs~\cite{rouard2023hybrid}& 8.48 & 7.94 & 5.72 & 7.93 & 7.52 \\
        BS RNN~\cite{luo2023music}  & 7.22 & 9.01 & 6.70 & 10.01 & 8.24 \\
        Hybrid Demucs~\cite{defossez2021hybrid} & 8.43 & 8.12 & 5.65 & 8.35 & 7.64 \\
        BS RoFormer~\cite{lu2024music} & 11.31 & 9.49 & 7.73 & 10.66 & 9.80 \\
        \midrule

        Demucs (s=10) + MSG & 6.07 & 5.78 & -- & 6.48 & 6.11 \\       
        
        \midrule
        BS-RF Det. & 9.25 & 11.57 & 7.99 & 10.55 & 9.84 \\
        BS-RF Det. $\times 2$ & 9.32 & 11.37 & 7.96 & 10.54 & 9.80 \\

        \midrule
        BS-RF + Diff. \scriptsize{($T=3, \sigma_{\text{max}}=0.002$)} & 10.12 & 12.10 & 8.21 & 10.94 & 10.34 \\
        \midrule
        BS-RF + CD \scriptsize{($T=1, \sigma_{\text{max}}=0.004$)} & 10.20 & 12.15 & \textbf{8.27} & \textbf{11.02} & \textbf{10.41} \\
        BS-RF + CD \scriptsize{($T=2, \sigma_{\text{max}}=0.003$)} & \textbf{10.24} & \textbf{12.17} & 8.24 & 10.95 & 10.40 \\


        \bottomrule
    \end{tabular}
}
\caption{\small\textbf{BS-RoFormer Experiments: Source Separation Results on MUSDB18.} The upper section presents \textbf{SI-SDR\(_\text{I}\)} results, while the lower section presents \textbf{SDR} results, comparing our BS-RF deterministic, diffusion, and CD models against baselines.} 
\label{tab:results_musdb}
\end{table}

\textbf{Consistency model.} Our CD model not only matches the Diffusion model at a single inference step but continues to improve with more steps. As observed in the last three rows of the table, CD with \(T=1\) nearly matches the Diffusion model in SI-SDR\(_\text{I}\) while slightly exceeding it in SDR. CD with \(T=2\) steps surpasses the Diffusion model by $\sim$0.8 dB in SI-SDR\(_\text{I}\) and $\sim$0.4 dB in SDR, demonstrating the ``student beating the teacher'' effect, which we attribute to the DSM auxiliary loss providing direct supervision from the target signal during distillation.
Finally, the best-performing CD model with \(T=4\) steps achieves a $\sim$2.9 dB SI-SDR\(_\text{I}\) gain and $\sim$1.05 dB SDR gain over the Deterministic model, and a dramatic $+6.5$ dB SI-SDR\(_\text{I}\) improvement over the strongest baseline, setting a new benchmark for MSS on Slakh2100. 

\textbf{SI-SDR\(_\text{I}\) evaluation.} During evaluation, we noticed unusually high SI-SDR\(_\text{I}\) scores for guitar and piano — not only in absolute terms, but also in model-to-model gains. Examining the formula (SI-SDR\(_\text{I}\) = SI-SDR$(\hat{s}, s)$ $-$ SI-SDR$(x_{\text{mix}}, s)$), we found that the second term measuring mixture–target alignment behaves poorly for silent (all-zero) stems: for an \emph{active} stem it is a moderately negative value ($\sim$$-$5 to $-$10 dB), but for a \emph{silent} stem it becomes a very large negative value whose magnitude is independent of separation quality and only depends on the numerical stability constant used in the implementation (on the order of $-$120\,dB in ours) — driving SI-SDR\(_\text{I}\) to an enormous value that is structurally incomparable across stems. This effect is most prominent for guitar and piano, which have low activity rates 55.04\% and 72.52\%, compared to bass 89.73\% and drums 97.08\%. SDR metric encounters a related problem: it becomes numerically undefined for all-zero stems, so \textit{museval} package filters them by design~\cite{stoter20182018}, making SDR score more reliable.
Following the same idea, for our BS-RoFormer experiments— where we are not tied to a specific baseline's evaluation pipeline — we exclude silent stems from SI-SDR\(_\text{I}\) evaluation. On Slakh2100 we preserve the MSDM pipeline for direct comparability. 

\textbf{Leakage and artifacts.} We hypothesized that the generative refinement would reduce source leakage and reconstruction artifacts. The SIR and SAR metrics offer supplementary evidence for this: SIR (interference reduction) improves from 18.1 dB (Det.) to 18.9 dB (Diff.) and 20.5 dB (CD $T=4$), while SAR (artifact reduction) improves from 12.2 dB to 12.6 dB and 13.2 dB respectively, suggesting a consistent reduction in both leakage and artifacts. We note, however, that SIR and SAR are part of the BSSEval family and are known to correlate poorly with human perception, particularly for generative models~\cite{jaffe2025bakeoff, bereuter2025reliable}; we therefore report them as supplementary indicators only.




\subsection{BS-RoFormer on MUSDB18}

Table~\ref{tab:results_musdb} reports SI-SDR\(_\text{I}\) and SDR results on our BS-RoFormer experiments with MUSDB18. The results show the following:

\textbf{Deterministic model.} The pre-trained BS-RoFormer we use as our Deterministic model (see Sec.~\ref{sec:bs_rf_arch}) matches the published BS-RoFormer numbers~\cite{lu2024music} in overall average SDR — which already outperforms all baselines — though per-instrument scores differ; both are included in the table for reference. For fair comparison, we only include baselines trained solely on MUSDB18.

\textbf{Diffusion model.} Adding the Diffusion model yields gains of $+0.48$\,dB SI-SDR\(_\text{I}\) and $+0.50$\,dB SDR over the Deterministic model. As in the U-Net experiments, BS-RF Det.$\times$2 serves as a control: stacking a second Deterministic model does not improve over the single Det., confirming the gains come from the generative component rather than added model capacity. We also evaluate MSG on MUSDB18 using the authors' released checkpoint on top of Demucs ($s=10$); the \textit{Other} stem is excluded as no checkpoint for it was provided. Similar to U-Net experiments, no objective improvement is observed.

\textbf{Consistency model.} The CD model at $T=1$ already surpasses the Diffusion model in both SI-SDR\(_\text{I}\) ($+0.13$\,dB) and SDR ($+0.07$\,dB), establishing a new state of the art on the MUSDB18 benchmark. Unlike the U-Net experiments, no additional quality improvement were observed with CD steps $T>1$ on this backbone.

\textbf{Leakage and artifacts.} SIR and SAR offer supplementary evidence for leakage and artifact reduction: SIR improves from 17.46\,dB (Det.) to 17.86\,dB (Diff., $+0.40$\,dB) and 17.83\,dB (CD $T=1$), while SAR improves from 10.65\,dB to 11.20\,dB (Diff., $+0.55$\,dB) and 11.29\,dB (CD $T=1$), suggesting a reduction in both leakage and artifacts, consistent with our U-Net experiment findings.

\textbf{Cross-experiment comparison.} While our BS-RoFormer experiments SI-SDR\(_\text{I}\) gains are smaller than in the U-Net experiments (due to the inflated score), SDR gains from the Diffusion step are comparable across both experiments ($+0.50$\,dB on MUSDB18 vs.\ $+0.45$\,dB on Slakh) — even though MUSDB18 is a $\sim$15$\times$ smaller dataset of acoustically complex real-world recordings and the backbone is already state-of-the-art. This demonstrates that our method generalises across backbone architectures, yielding reliable improvements regardless of the underlying separator.

\begin{table}[t]
    \centering    
    \resizebox{\linewidth}{!}{%
    \begin{tabular}{c c c c c}
            \toprule
        Model                                                   & Inference Time (ms)            & \# of parameters & RTF \\
        \midrule
        \multicolumn{4}{c}{\textit{Slakh2100 (11.9s segment, 22.05 kHz)}} \\
        \midrule
        Demucs (small, s=0) \cite{defossez2019music}\cite{mariani2024multisource}       & 111                          & 40M          & 0.009 \\
        Demucs + Gibbs (512) \cite{manilow2022improving}\cite{mariani2024multisource}  & $111 \times 512 = 56{,}832$  & ${\sim}$40M   & 4.776 \\
        Demucs + Gibbs (256) \cite{manilow2022improving}\cite{mariani2024multisource}  & $111 \times 256 = 28{,}416$  & ${\sim}$40M   & 2.388\\
        ISDM  {\scriptsize ($T \times R = 150 \times 2$)}~\cite{mariani2024multisource}         & $4600 \times 4 = 18{,}400$    & 405M $\times 4$   & 1.546\\
        MSDM  {\scriptsize ($T \times R = 150 \times 2$)}~\cite{mariani2024multisource}          & 4600                         & 405M              & 0.386\\
        \midrule
        U-Net Det.                                            & $28.5 \times 4 = 114$        & 405M             & 0.009\\
        U-Net Det. $\times 2$                                 & $28.5 \times 2 \times 4 = 228$ & 405M $\times 2$  & 0.019\\
        U-Net Det. + MSG                                      & $(28.5 + 1080) \times 4 = 4434$ & 405M + 90M $\times 4$ & 0.369\\
        U-Net + Diff. \scriptsize({$T \times R = 2 \times 2$})  & $(28.5 + 4 \times 28.5) \times 4 = 570$ & 405M $\times 2$  & 0.048\\
        U-Net + CD \scriptsize({$T=1$})                        & $(28.5 + 1 \times 28.5) \times 4 = 228$ & 405M $\times 2$  & 0.019\\
        U-Net + CD \scriptsize({$T=2$})                        & $(28.5 + 2 \times 28.5) \times 4 = 342$ & 405M $\times 2$  & 0.029\\
        U-Net + CD \scriptsize({$T=4$})                        & $(28.5 + 4 \times 28.5) \times 4 = 570$ & 405M $\times 2$  & 0.048\\
        \midrule
        \multicolumn{4}{c}{\textit{MUSDB18 (11s segment, 44.1 kHz)}} \\
        \midrule
        Demucs ($s=0$)  & 105                          & 265M        & 0.010 \\
        Demucs ($s=10$)  & 1087                        & 265M        & 0.099 \\
        Demucs ($s=10$) + MSG & $1087 + 4 \times 1050 = 5287$ & 265M + 90M $\times 4$  & 0.481 \\
        BS-RF Det.                                            & 231                          & 131M           & 0.021 \\
        BS-RF + Diff. \scriptsize{($T=3$)}             & $231 + 3 \times 265 \times 4 = 3411$ & 131M + 180M    & 0.310 \\
        BS-RF + CD \scriptsize{($T=1$)}                & $231 + 1 \times 265 \times 4 = 1291$ & 131M + 180M    & 0.117 \\
        BS-RF + CD \scriptsize{($T=2$)}                & $231 + 2 \times 265 \times 4 = 2351$ & 131M + 180M    & 0.214 \\

        \bottomrule
    \end{tabular}
    }
        \caption{\small\textbf{Inference Speed and Parameter Count.} Inference time, parameter count, and RTF comparison. Upper section: U-Net on Slakh2100 (11.9s) and baselines; lower section: BS-RoFormer on MUSDB18 (11s) and baselines. For models that process one stem at a time (U-Net, ISDM, MSG, BS-RF Diff/CD), times are multiplied by four to reflect full separation of all stems.}
    \label{tab:inference_time}
\end{table}

\subsection{Inference Speed and Parameter Count}

Beyond quality, inference efficiency is a key motivation of our work. Table~\ref{tab:inference_time} reports inference time, parameter count, and Real-Time Factor (RTF) for both experimental tracks. Since models that process one stem at a time (all of U-Net, ISDM, MSG, BS-RF Diff/CD) must be run four times for full separation, all reported times are scaled accordingly.

Our U-Net Det.\ runs at 114\,ms (RTF\,0.009), comparable to the lightweight Demucs ($s=0$, no shift trick) used in~\cite{manilow2022improving} at 111\,ms. All generative baselines are substantially slower: MSDM requires 4.6\,s, Demucs+Gibbs with 256 steps 56\,s, and ISDM 18.4\,s (one stem at a time, run $\times4$). Our U-Net + Diff.\ configuration (570\,ms, RTF\,0.048) is more than $30\times$ faster than MSDM while achieving better quality. CD models compress this further: U-Net + CD ($T=1$) matches the parameter-doubled Det.\ $\times 2$ at 228\,ms (RTF\,0.019), and even the $T=4$ setting equals the diffusion model at 570\,ms — confirming that CD provides quality gains without additional cost over the diffusion model.

In our BS-RoFormer experiments, BS-RF Det.\ outputs all four stems simultaneously; however, the Diff.\ and CD models process one stem at a time, making them proportionally slower — owing also to the added noise embedding and adapter layers, which slightly increase model size and per-pass cost (231\,ms $\rightarrow$ 265\,ms). BS-RF + Diff.\ is the slowest, while BS-RF + CD ($T=1$) and ($T=2$) take 1\,291\,ms (RTF\,0.117) and 2\,351\,ms (RTF\,0.214) respectively, both well within practical deployment ranges while delivering meaningful quality gains. 
While batching all four stems into a single forward pass is theoretically possible, in practice it yields no acceleration: the transformer architecture with large time-frequency sequence length (${\sim}$283k tokens at 44.1\,kHz) fully saturate GPU compute already at batch size 1, as confirmed on an A6000. Comparing to other baselines, Demucs with the recommended shift trick ($s=10$) takes 1\,087\,ms — nearly $5\times$ slower than BS-RF Det.\ — and adding MSG makes that deterministic-generative hybrid substantially slower than ours, with our quality being far superior.

\section{Conclusion}
We proposed a refinement framework for music source separation in which a diffusion or consistency model post-processes the output of a deterministic separator, improving separation quality. Across two experimental tracks — a custom U-Net on Slakh2100 and a state-of-the-art BS-RoFormer on MUSDB18 — our diffusion model consistently improves over the deterministic backbone across standard benchmark metrics (SI-SDR\(_\text{I}\) and SDR). Consistency distillation then recovers most of these gains in a single step, with inference time comparable to running the deterministic model twice, and improves further with additional steps. The framework is architecture-agnostic: the same design applies equally to a custom U-Net and a state-of-the-art transformer, suggesting that any strong separator — present or future — can serve as a backbone and benefit from this refinement paradigm.

\section{Acknowledgements}
Part of this work was carried out during an internship at Bose Corporation. The authors gratefully acknowledge support from the Institute for Research and Coordination in Acoustics and Music (IRCAM) under Project REACH: Raising Co-creativity in Cyber-Human Musicianship, funded by the European Research Council (ERC) under the European Union's Horizon 2020 research and innovation programme (Grant Agreement No.\ 883313).


\bibliography{DiCoSe}
\end{document}